\documentclass[letterpaper, 10 pt, conference]{ieeeconf}
\IEEEoverridecommandlockouts 

\usepackage{cite}
\usepackage{amsmath}
\usepackage{amssymb}
\usepackage{bm}
\let\labelindent\relax
\usepackage{enumitem}
\usepackage{algorithm}
\usepackage{algpseudocode}
\usepackage{graphicx}
\usepackage[mathscr]{eucal}
\usepackage{xcolor}
\usepackage[hidelinks]{hyperref}
\graphicspath{{figures/}}

\newtheorem{theorem}{Theorem}
\newtheorem{lemma}{Lemma}
\newtheorem{assumption}{Assumption}
\newtheorem{definition}{Definition}
\newtheorem{remark}{Remark}
\newtheorem{corollary}{Corollary}
\newtheorem{proposition}{Proposition}

\DeclareMathOperator{\Tr}{Tr}

\DeclareRobustCommand{\trace}[1]{\Tr\left(#1\right)}

\definecolor{orange}{rgb}{1,0.65,0.2}
\definecolor{darkgreen}{rgb}{0.0, 0.4, 0.0}

\newcommand{\mat}[1]{\begin{bmatrix} #1 \end{bmatrix}}

\newcommand{\R}{\mathbb{R}}

\newcommand{\cK}{\mathscr{K}}

\newcommand{\cP}{\mathscr{P}}

\newcommand{\sC}{\mathcal{C}}
\newcommand{\sD}{\mathcal{D}}

\newcommand{\sT}{\mathcal{T}}
\newcommand{\sU}{\mathcal{U}}
\newcommand{\sV}{\mathcal{V}}

\newcommand{\sX}{\mathcal{X}}

\begin{document}

\title{Designing Control Barrier Functions  Using a Dynamic Backup Policy} 
\author{Victor Freire, and Marco M. Nicotra
\thanks{This research was supported by the NSF-CMMI Award \#2411667.}
\thanks{The authors are with the Department of Electrical, Computer \& Energy Engineering, University of Co\-lorado,  Boulder, CO 80309 USA (email: vifr9883@colorado.edu; marco.nicotra@colorado.edu).} \thanks{\emph{(Corresponding author: Victor Freire.)}}}

\maketitle

\begin{abstract}
  This paper presents a systematic approach to construct control barrier functions for nonlinear control affine systems subject to arbitrary state and input constraints. Taking inspiration from the reference governor literature, the proposed method defines a family of backup policies, parametrized by the equilibrium manifold of the system. The control barrier function is defined on the augmented state-and-reference space: given a state-reference pair, the approach quantifies the distance to constraint violation at any time in the future. The proposed method is applied to an inverted pendulum on cart.
\end{abstract}


\section{Introduction}
Control barrier functions (CBFs) are powerful tools to design controllers for safety-critical systems. However, they are difficult to synthesize for general systems under  arbitrary state and input constraints. In many applications, practitioners resort to candidate CBFs because they are easier to design and achieve good performance with tuning and the use of slack variables. However, they lack the rigorous safety guarantees of valid CBFs.

Although the modern definition of CBFs was stated in \cite{ames2016control}, their systematic design remains an open problem. Recent techniques include the following: \cite{harms2024neural} uses machine learning to design and implement CBFs; 
\cite{dai2024verification} uses sum-of-squares to design polynomial CBFs;
\cite{jang2024safe} uses control Lyapunov functions to design families of CBFs; 
\cite{lindemann2024learning} uses expert demonstrations to design robust CBFs; 
\cite{shakhesi2025counterexample} partitions the state-space to search for counterexamples to guide the construction of CBFs;
\cite{choi2021robust}, proposes robust control barrier-value functions, which unify the Hamilton--Jacobi reachability and CBF methods.
Most relevant to this paper, \cite{chen2021backup} and \cite{van2024disturbance} propose backup CBFs, which leverage a pre-specified control policy to maintain safety guarantees.

Prior work by the authors showed that it is possible to design proper CBFs using tools from the reference governor literature. Specifically, \cite{freire2026using} proved that dynamic safety margins (DSMs) are CBFs in the augmented state-reference space. This led to the definition of Lyapunov-based DSM-CBFs in \cite{freire2026using} and passivity-based DSM-CBFs in \cite{freire2025designing}. In this paper, we extend the trajectory-based DSMs proposed in \cite{nicotra2018explicit} to construct trajectory-based DSM-CBFs. Although the resulting approach is conceptually similar to the backup CBFs detailed in \cite{chen2021backup} and \cite{van2024disturbance}, this paper features several key contributions: (a) the backup policy is parametrized via the equilibrium manifold of the system, (b) the backup policy is dynamic because its parametrization is time-varying (c) under mild assumptions, it is shown that the underlying sensitivity matrices are asymptotically stable, and (d) the potential nonsmoothness of the resulting CBF is rigorously addressed using Clarke generalized Jacobians. Open source code of the proposed numerical validations can be found in our GitHub repository\footnote{\texttt{https://github.com/ROCC-Lab-CU-Boulder/traj-DSM-CBF}} and can be easily modified to other constrained systems.


\section{Preliminaries} \label{sec:preliminaries}
Consider a control affine system
\begin{equation} \label{eq:sys}
    \dot{\bm{x}} = f(\bm{x}) + g(\bm{x}) \bm{u},
\end{equation}
subject to state and input constraints $\bm x\in\sX$ and $\bm u\in\sU$.
\begin{assumption}
    The functions $f\!:\!\R^n \!\to\! \R^n$ and $g\!:\!\R^n \!\to\! \R^{n\times m}$ are continuously differentiable.
\end{assumption}
\begin{assumption}
    The sets $\sX \subset \R^n$ and $\sU \subset \R^m$ are closed.
\end{assumption}
Our objective is to provide a simple yet rigorous approach to enforce constraints using control barrier functions. To this end, we include a standard assumption from the reference governor (RG) literature that applies to any system that admits a connected path of stabilizable equilibrium points. As noted in \cite{garone2017reference}, this encompasses a wide range of meaningful systems and applications.
\begin{assumption}\label{ass:equilibrium}
The control affine system \eqref{eq:sys} admits an equilibrium manifold parameterized by $\bm{v} \in \R^l$. Specifically, there exist continuous functions $\bar{x}:\R^l \to \R^n$ and $\bar{u} : \R^l \to \R^m$ such that
\begin{equation}
    \forall \bm{v} \in \R^l, \quad f\big(\bar{x}(\bm{v})\big) + g\big(\bar{x}(\bm{v})\big) \bar{u}(\bm{v}) = 0.
\end{equation}
Moreover, there exists a continuously differentiable control policy $\pi: \R^n \times \R^l \to \R^m$ such that $\bar{x}(\bm{v})$ is an asymptotically stable equilibrium point with an open region of attraction $\sD_{\bm{v}} \subset \R^n$.
\end{assumption}
Assumption \ref{ass:equilibrium} requires an understanding of the equilibrium points of the system and the ability to design a (local) stabilizing control law. 
The following subsections summarize existing results from the CBF and RG literatures.

\subsection{Control Barrier Functions}
Control barrier functions can be used to synthesize safety filters that modify a nominal control law $\kappa(\bm x)$ as little as necessary to enforce constraints.
\begin{definition}[\cite{freire2026using}]
  A continuously differentiable function $h: \sD \to \R^{p}$ is a \emph{control barrier function} (CBF) if there exists $\alpha \in \cK$ for which, $\forall \bm{x} \in \sC$, $\exists \bm{u} \in \sU$, such that 
  \begin{equation}
    \min_{i \in \{1,\ldots,p\}} \left[L_fh_i(\bm{x}) + L_gh_i(\bm{x}) \bm{u} + \alpha\big(h_i(\bm{x})\big)\right] \geq 0, 
  \end{equation}
  where $\sC = \{\bm{x} \in \sD \mid h(\bm{x}) \geq 0\}$, and $L_fh$ and $L_gh$ are the Lie derivatives of $h$ along $f$ and $g$, respectively.
\end{definition}
This definition is a generalization of the CBFs presented in \cite{ames2016control} to vector-valued functions that leverages the control-sharing property detailed in \cite{xu2018constrained}.
If $\sC \subset \sX$, the solution to the CBF-based safety filter
\begin{align} \label{eq:cbf-sf}
    \min_{\bm{u} \in \sU} \quad &\|\bm{u} - \kappa(\bm{x})\|^2 \\
    \text{s.t.} \quad & L_fh_i(\bm{x}) + L_gh_i(\bm{x}) \bm{u} \geq -\alpha\big(h_i(\bm{x})\big), \; i = 1,\ldots,p \nonumber
\end{align}
can be used as a control policy that approximates $\kappa(\bm{x})$ as much as possible while also enforcing constraints.

Although simple to implement, the main drawback of CBFs is that finding a suitable function $h(\bm{x})$ is generally challenging. However, in \cite{freire2026using}, we showed that dynamic safety margins are CBFs.

\subsection{Dynamic Safety Margins}
Under Assumption \ref{ass:equilibrium}, let $\pi(\bm{x}, \bm{v})$ be a prestabilizing control policy. Given the prestabilized dynamics
\begin{equation}\label{eq:sys-prestab}
    \dot{\bm{x}} = f_{\pi}(\bm{x}, \bm{v}) \triangleq f(\bm{x}) + g(\bm{x}) \pi(\bm{x}, \bm{v}),
\end{equation} 
dynamic safety margins quantify the risk of constraint violation at any time in the future, should the virtual reference $\bm{v}$ remain constant.

In the context of this paper, it is convenient to generalize sets of interest to the augmented state-reference space $\R^n \times \R^l$. Notably, we define the augmented constraint set
\begin{equation}
  \tilde{\sX} \triangleq \{(\bm{x}, \bm{v}) \in \R^n \times \R^l \mid \bm{x} \in \sX_{\bm{v}},~  \bm{v} \in \sV\},
\end{equation}
where $\sV = \{\bm{v} \in \R^l \mid \bar{x}(\bm{v}) \in \sX,~ \bar{u}(\bm{v}) \in \sU\}$ is the set of steady-state admissible references and
\begin{equation}\label{eq:Xv_base}
    \sX_{\bm{v}} = \{\bm{x}\in\R^n\mid x \in \sX ,~ \pi(\bm{x}, \bm{v}) \in \sU\}
\end{equation}
is the reference-dependent set of state constraints. Note that $\tilde{\sX}$ is closed in $\R^n \times \R^l$. The following technical assumption ensures that the prestabilizing policy has a well-behaved region of attraction.

\begin{assumption} \label{ass:stab-large}
    The augmented region of attraction 
    \begin{equation}
        \tilde{\sD} \triangleq \{(\bm{x}, \bm{v}) \in \R^n \times \R^l \mid \bm{x} \in \sD_{\bm{v}}\},
    \end{equation}
    is open in $\R^n \times \R^l$. Moreover, the augmented constraint set satisfies $\tilde{\sX} \subset \tilde{\sD}$.
\end{assumption}

\begin{remark}
    With no loss of generality, $\tilde{\sX} \subset \tilde{\sD}$ can always be enforced by introducing state constraints that capture the stability requirements of the prestabilizing controller.
\end{remark}

We are now ready to provide a slightly stronger definition of DSMs that requires smoothness and invariance over continuity and returnability. For the original definition, the reader is referred to \cite{nicotra2018explicit}.

\begin{definition} \label{def:dsm}
  Let $\Delta: \tilde{\sD} \to \R^{p}$ be continuously differentiable and define the set $\tilde{\sC} = \{(\bm{x}, \bm{v}) \in \tilde{\sD} \mid \Delta(\bm{x}, \bm{v}) \geq 0\}$. The function $\Delta$ is called a \emph{dynamic safety margin} (DSM) if the following conditions hold
  \begin{subequations}
    \begin{align}
      &\tilde{\sC} \subset \tilde{\sX}, \label{eq:dsm-safe}\\    
      &\tilde{\sC} ~ \text{is compact in} ~ \R^n \times \R^l, \label{eq:dsm-compact} \\
      &\Delta_i(\bm{x}, \bm{v}) = 0 ~ \implies ~ \frac{\partial \Delta_i(\bm{x}, \bm{v})}{\partial \bm{x}}f_{\pi}(\bm{x}, \bm{v}) \geq 0, \label{eq:dsm-invariant}
    \end{align}
  \end{subequations}
  where the last property holds $\forall i \in \{1,\ldots,p\}$.
\end{definition}

\subsection{DSM-based CBFs}
The following theorem states that DSMs are CBFs for an augmented system consisting of the concatenation of the state $\bm{x}$ and the virtual reference $\bm{v}$
\begin{equation} \label{eq:sys-aug}
  \mat{\dot{\bm{x}} \\ \dot{\bm{v}}} = \mat{f(\bm{x}) + g(\bm{x})\bm{u} \\ \bm{w}},
\end{equation}
with augmented input $(\bm{u}, \bm{w}) \in \R^m \times \R^l$.
\begin{theorem}[\cite{freire2026using}] \label{thm:dsm-cbf}
  If $\Delta: \tilde{\sD} \to \R^{p}$ is a DSM, then $\Delta$ is a CBF for the augmented system \eqref{eq:sys-aug}.
\end{theorem}
In \cite{freire2026using}, Theorem \ref{thm:dsm-cbf} is leveraged to construct CBFs using Lyapunov-based DSMs. In this paper, we construct CBFs using the trajectory-based DSM
\begin{equation}
  \Delta_i(\bm{x}, \bm{v}) = \inf_{\tau \in [0,\infty)} ~ c_i\big(\Phi(\tau, \bm{x}, \bm{v}), \bm{v} \big),
\end{equation}
where $c:\R^n \times \R^l \to \R^{p}$ is a continuously differentiable function describing the reference-dependent set of state constraints (i.e., $\sX_{\bm{v}} = \{\bm{x} \in \R^n \mid c(\bm{x}, \bm{v}) \geq 0\}$) and $\Phi(\tau, \bm{x}, \bm{v})$ is the solution to the prestabilized dynamics \eqref{eq:sys-prestab}, given the initial condition $\bm{x}(0) = \bm{x}$ and a constant reference $\bm{v}$. The following section computes the flow sensitivity Jacobians $\partial \Phi/\partial\bm{x}$ and $\partial \Phi/\partial\bm{v}$, which will appear in the Lie derivatives of $\Delta(\bm{x}, \bm{v})$.

\section{Sensitivity Analysis of the Prestabilized Dynamics} \label{sec:sensitivity}
When there is no ambiguity, we denote by $\Phi(\tau) = \Phi(\tau, \bm{x}, \bm{v})$ the solution to the initial value problem
\begin{equation}\label{eq:Phi-ivp}
    \frac{\partial \Phi(\tau) }{\partial \tau} = f_{\pi}\big(\Phi(\tau), \bm{v}\big), \qquad \Phi(0) = \bm{x}.
\end{equation}
By the fundamental theorem of calculus, 
\begin{equation} \label{eq:Phi}
    \Phi(\tau) = \bm{x} + \int_{0}^{\tau} f_{\pi}\big(\Phi(\sigma), \bm{v}\big) ~ \mathrm{d}\sigma.
\end{equation}
Its derivative with respect to the initial condition $\bm{x}$ is
\begin{equation}\label{eq:calc_Px}
  \frac{\partial \Phi (\tau)}{\partial \bm{x}} = I_n +
  \int_{0}^{\tau} \frac{\partial f_{\pi}\big(\Phi(\sigma),\bm{v}\big)}{\partial \bm{x}} \frac{\partial \Phi(\sigma) }{\partial
  \bm{x}} ~ \mathrm{d}\sigma.
\end{equation}
Let us define the state sensitivity Jacobian
\begin{equation}
  S_{\bm{x}}(\tau) = \frac{\partial \Phi (\tau)} {\partial \bm{x}}.
\end{equation}
Then, it follows from \eqref{eq:calc_Px} that $S_{\bm{x}}(\tau)$ can be obtained by solving the initial value problem
\begin{equation} \label{eq:Sx}
  \frac{\partial S_{\bm{x}}(\tau)} {\partial \tau}  =  \frac{\partial f_{\pi}\left(\Phi(\tau), \bm{v}\right)}
  {\partial \bm{x}}
  S_{\bm{x}}(\tau), \qquad S_{\bm{x}}(0) = I_n.
\end{equation}

In a similar manner, we note that the derivative of \eqref{eq:Phi} with respect to the virtual reference $\bm{v}$ is
\begin{equation*}
  \frac{\partial \Phi(\tau)}{\partial \bm{v}} = \int_{0}^{\tau} \frac{\partial f_{\pi}\big(\Phi(\sigma),\bm{v}\big)}{\partial \bm{x}} \frac{\partial \Phi(\sigma)}{\partial \bm{v}} + \frac{\partial f_{\pi}\big(\Phi(\sigma),\bm{v}\big)}{\partial \bm{v}}  ~ \mathrm{d}\sigma.
\end{equation*}
By defining the reference sensitivity Jacobian
\begin{equation}
  S_{\bm{v}}(\tau) = \frac{\partial \Phi(\tau)}{\partial \bm{v}},
\end{equation}
it follows that we can compute $S_{\bm{v}}(\tau)$ as the solution to 
\begin{equation} \label{eq:Sv}
  \frac{\partial S_{\bm{v}}(\tau)}{\partial \tau} = \frac{\partial f_{\pi}\big(\Phi(\tau),\bm{v}\big)}{\partial \bm{x}} S_{\bm{v}}(\tau) + \frac{\partial f_{\pi}\big(\Phi(\tau), \bm{v}\big)}{\partial \bm{v}},
\end{equation}
subject to the initial condition $S_{\bm{v}}(0) = 0_{n\times l}$.

The following proposition states that, under mild assumptions, the flow sensitivity Jacobians $S_{\bm{x}}(\tau)$ and $S_{\bm{v}}(\tau)$ converge to values that are known a-priori.
\begin{proposition}
  If $\frac{\partial f_{\pi}(\bar{x}(\bm{v}), \bm{v})}{\partial \bm{x}}$ is Hurwitz, then $\forall\bm{x} \in \sD_{\bm{v}}$, the solutions to \eqref{eq:Sx} and \eqref{eq:Sv} satisfy
  \begin{subequations}\label{eq:S-asymptotes}
    \begin{align}
     \lim_{\tau\to\infty} S_{\bm{x}}(\tau) =& \ 0,\\
     \lim_{\tau\to\infty} S_{\bm{v}}(\tau) =& - \frac{\partial f_{\pi} \big(\bar{x}(\bm{v}), \bm{v}\big)}{\partial \bm{x}}^{-1}\ \frac{\partial f_{\pi}\big(\bar{x}(\bm{v}), \bm{v}\big)}{\partial \bm{v}}.
    \end{align}
  \end{subequations}
\end{proposition}
\begin{proof}
  Let
  \begin{equation*}
    A_{\pi}(\tau) = \frac{\partial f_{\pi}\big(\Phi(\tau), \bm{v}\big)}{\partial \bm{x}}, \quad \bar{A}_{\pi} = \frac{\partial f_{\pi}\big(\bar{x}(\bm{v}), \bm{v}\big)}{\partial \bm{x}} \prec 0.
  \end{equation*}
  Since $A_{\pi}(\tau)$ is continuous and $\lim_{\tau \to \infty}A_{\pi}(\tau) = \bar{A}_{\pi}$, the solution to the initial value problem \eqref{eq:Sx}, namely $S_{\bm{x}}(\tau)$, exists and is continuous on $[0,\infty)$. Let $\bar{P} \succ 0$ be the unique solution to the Lyapunov equation $\bar{A}_{\pi}^\top \bar{P} + \bar{P} \bar{A}_{\pi} + I_n = 0$. Consider the Lyapunov candidate function 
  \begin{equation}\label{eq:Lyap}
    V(X) = \trace{X^\top \bar{P} X}.
  \end{equation}
  Its derivative along the state sensitivity Jacobian flow is
  \begin{equation}
    \frac{\partial V\big(S_{\bm{x}}(\tau) \big)}{\partial \tau} =- \trace{S_{\bm x}(\tau)^\top  Q(\tau) S_{\bm{x}}(\tau)}, 
  \end{equation}
where $Q(\tau) = - A_{\pi}(\tau)^\top \bar{P} - \bar{P} A_{\pi}(\tau)$ is symmetric. Since $\bm{x} \in \sD_{\bm{v}}$, Assumption \ref{ass:equilibrium} ensures $\lim_{\tau \to \infty} \Phi(\tau) = \bar x(\bm v)$. Therefore, $\lim_{\tau \to \infty} Q(\tau) = I_n$ and there exists $T > 0$ such that $Q(\tau) \succ I_n/2$, $\forall \tau \geq T$. Therefore,
  \begin{equation}
    \frac{\partial V\big(S_{\bm{x}}(\tau) \big)}{\partial \tau} < - \frac{1}{2} \|S_{\bm{x}}(\tau)\|^2_F,\quad \forall \tau \geq T,
  \end{equation}
  where $\|X\|_F=\sqrt{\trace{X^\top X}}$ is the Frobenius norm of $X$. Since $V\big(S_{\bm{x}}(T)\big)$ is bounded and $\frac{\partial V(S_{\bm{x}}(\tau))}{\partial \tau} < 0,~\forall \tau\geq T$, we have $\lim_{\tau \to \infty} V\big(S_{\bm{x}}(\tau)\big) = 0$. Since $\bar{P} \succ 0$, we conclude that $\lim_{\tau \to \infty} S_{\bm x}(\tau) = 0$. For the second claim, let
  \begin{equation*}
    B_{\pi}(\tau) = \frac{\partial f_{\pi}\big(\Phi(\tau), \bm{v}\big)}{\partial \bm{v}},
    \quad \bar{B}_{\pi} = \frac{\partial f_{\pi}\big(\bar{x}(\bm{v}),
    \bm{v}\big)}{\partial \bm{v}},
  \end{equation*}
  and define the error matrix $E(\tau) = S_{\bm{v}}(\tau) + \bar{A}_{\pi}^{-1} \bar{B}_{\pi}$. Then,
  \begin{align*}
    \frac{\partial E(\tau)}{\partial \tau} &= A_{\pi}(\tau)E(\tau) - A_{\pi}(\tau)
    \bar{A}_{\pi}^{-1} \bar{B}_{\pi} + B_{\pi}(\tau) \\ &= A_{\pi}(\tau) E(\tau) + D(\tau),
  \end{align*}
  where the disturbance term $D(\tau)\! =\! B_{\pi}(\tau) - A_{\pi}(\tau) \bar{A}_{\pi}^{-1} \bar{B}_{\pi}$ satisfies $\lim_{\tau\to\infty}D(\tau)=0$. As before, it follows from continuity of $A_\pi(\tau)$ and $B_\pi(\tau)$ that $E(\tau)$ exists and is unique $\forall \tau\in[0,\infty)$. Evaluating the derivative of \eqref{eq:Lyap} along the error term $E(\tau)$, we obtain
  \begin{equation*}
    \frac{\partial V\big(E(\tau)\big)}{\partial \tau} \!=\! 2\trace{E(\tau)^\top \bar{P} D(\tau)} -\trace{E(\tau)^\top Q(\tau) E(\tau)},
  \end{equation*}
  As before, we can show that, $\forall \tau \geq T$,
  \begin{equation}
    \frac{\partial V\big(E(\tau)\big)}{\partial \tau}
    < -\frac{1}{2} \|E(\tau)\|^2_F +  2\|E(\tau)
    \|_F\|\bar{P}D(\tau)\|_F,
  \end{equation}
  where we have used the Cauchy--Schwartz inequality. This is sufficient to show that, for all $\tau \geq T$, $V$ is an ISS Lyapunov function with respect to the
  disturbance $D(\tau)$. Since $V\big(E(T)\big)$ is bounded and $D(\tau)$ is time-vanishing, we conclude $\lim_{\tau \to \infty} V\big(E(\tau)\big) = 0$. Since
  $\bar{P} \succ 0$, we have $\lim_{\tau \to
  \infty} E(\tau) = 0$, thus $\lim_{\tau \to \infty} S_{\bm{v}}(\tau) =
  -\bar{A}_{\pi}^{-1} \bar{B}_{\pi}$.\smallskip
\end{proof}
\begin{remark}
Since the equilibrium point $\bar{x}(\bm{v})$ is asymptotically stable by construction, the only exception to \eqref{eq:S-asymptotes} is when the Lyapunov indirect method is inconclusive (e.g. $f_\pi(x, v)=-(x-v)^2$). As such, $\pi$ typically ensures the convergence of the flow sensitivity Jacobians.
\end{remark}

The following lemma shows how the state sensitivity Jacobian $S_{\bm{x}}(\tau)$ correlates a change in the present state $\dot{\bm{x}}$ to a change in the prestabilized dynamics at a future predicted state $\Phi(\tau)$.
\begin{lemma} \label{lem:sensitivity-property}
  For all $\bm{x} \in \R^n$, $\bm{v} \in \R^l$, $\tau \geq 0$,
  \begin{equation}
    S_{\bm{x}}(\tau)f_{\pi}(\bm{x}, \bm{v}) = f_{\pi}\big(\Phi(\tau), \bm{v}\big).
  \end{equation}
\end{lemma}
\begin{proof}
  Given $\bm{x} \in \R^n$ and $\bm{v} \in \R^l$, define $A_{\pi}(\tau) \triangleq \frac{\partial f_{\pi}\big(\Phi(\tau), \bm{v}\big)}{\partial \bm{x}}$, $L(\tau) \triangleq S_{\bm{x}}(\tau)f_{\pi}(\bm{x}, \bm{v})$, $R(\tau) \triangleq f_{\pi}\big(\Phi(\tau), \bm{v}\big)$.
  Note that $\frac{\partial L(\tau)}{\partial \tau} = \frac{\partial S_{\bm{x}}(\tau)}{\partial \tau} f_{\pi}(\bm{x}, \bm{v}) = A_{\pi}(\tau)L(\tau)$, and that $\frac{\partial R(\tau)}{\partial \tau} = \frac{\partial f_{\pi}\big(\Phi(\tau),\bm{v}\big)}{\partial \bm{x}} \frac{\partial \Phi(\tau)}{\partial \tau} = A_{\pi}(\tau)R(\tau)$.
  Finally, note that $L(0) = f_{\pi}(\bm{x}, \bm{v}) = R(0)$. Since $L(\tau)$ and $R(\tau)$ are both subject to the same linear ordinary differential equation with the same initial conditions, we conclude $R(\tau) \equiv L(\tau)$.
\end{proof}

\section{Trajectory Based DSM-CBF} \label{sec:traj-dsm}
In this section, we show how to construct CBFs starting from the trajectory-based DSMs featured in \cite{nicotra2018explicit}.
The following theorem constructs a trajectory-based DSM for the case of a single constraint $p=1$. This result trivially generalizes to vector-valued constraints $c:\R^n \times \R^l \to \R^{p}$.

\begin{theorem} \label{thm:traj-dsm}
  Assume $\sX \subset \R^n$ and $\sV \subset \R^l$ are compact. If the function $\Delta:\tilde{\sD} \to \R$, defined as
  \begin{equation} \label{eq:traj-dsm}
    \Delta(\bm{x}, \bm{v}) = \inf_{\tau \in [0,\infty)} ~ c\big(\Phi(\tau), \bm{v}   \big),
  \end{equation}
  is continuously differentiable, then it is a DSM.
\end{theorem}
\begin{proof}
  To prove property \eqref{eq:dsm-safe}, let $(\bm{x}, \bm{v}) \in \tilde{\sC}$. Then, we have $c(\bm{x}, \bm{v}) = c\big(\Phi(0), \bm{v}\big) \geq \Delta(\bm{x}, \bm{v})  \geq 0$, which implies $\bm{x} \in \sX_{\bm{v}}$. Also, $\lim_{\tau \to \infty} c\big(\Phi(\tau), \bm{v}\big) = c\left(\lim_{\tau \to \infty} \Phi(\tau), \bm{v}\right) = c\big(\bar{x}(\bm{v}), \bm{v}\big) \geq 0$, which implies $\bar{x}(\bm{v}) \in \sX_{\bm{v}}$ and $\bm{v} \in \sV$.

  To prove property \eqref{eq:dsm-compact}, note that $\tilde{\sC}$ is bounded since $\tilde{\sC} \subset \sX \times \sV$. Thus, we only need to show that $\tilde{\sC}$ is closed in $\R^n \times \R^l$. This property follows by continuity of $\Delta(\bm{x}, \bm{v})$ and the fact that, by Assumption \ref{ass:stab-large}, $\tilde{\sC} \subset \tilde{\sX} \subset \tilde{\sD}$, where $\tilde{\sX}$ is closed in $\R^n \times \R^l$.

  To prove property \eqref{eq:dsm-invariant}, pick $(\bm{x}, \bm{v}) \in \tilde{\sD}$ such that $\Delta(\bm{x}, \bm{v}) = 0$ and, for a contradiction, assume 
  \begin{equation*}
    \frac{\partial \Delta(\bm{x}, \bm{v})}{\partial \bm{x}} f_{\pi}(\bm{x}, \bm{v}) < 0.
  \end{equation*}
  Define the function $\xi:t \mapsto \Delta\big(\Phi(t), \bm{v}\big)$, which is
  continuously differentiable and satisfies
  \begin{align*}
    \xi(0) &= \Delta\big(\Phi(0), \bm{v}\big) = \Delta(\bm{x}, \bm{v}) = 0,\\
    \dot{\xi}(0) &= \frac{\partial \Delta\big(\Phi(0), \bm{v}\big)}{\partial \bm{x}}
    \dot{\Phi}(0) = \frac{\partial \Delta(\bm{x}, \bm{v})}{\partial \bm{x}} f_{\pi}(\bm{x},
    \bm{v}) < 0.
  \end{align*}
  By continuity of $\xi$, there exists $\epsilon > 0$ such that $\xi(\epsilon) = \Delta\big(\Phi(\epsilon), \bm{v}\big) < 0$. Finally, we obtain the contradiction $\Delta\big(\Phi(\epsilon), \bm{v}\big) = \inf_{\sigma \in [0,\infty)} c\left(\Phi\big(\sigma, \Phi(\epsilon), \bm{v}\big),\bm{v}\right) = \inf_{\sigma \in [0,\infty)} c\left(\Phi(\sigma \!+\! \epsilon, \bm{x}, \bm{v}),\bm{v}\right) \!\geq\! \inf_{\sigma \in [-\epsilon, \infty)} c(\Phi(\sigma + \epsilon), \bm{v}) = \inf_{\tau \in [0, \infty)}  c(\Phi(\tau), \bm{v}) = \Delta(\bm{x}, \bm{v}) = 0$.

\end{proof}

By Theorem \ref{thm:dsm-cbf}, the DSM $\Delta$ given in \eqref{eq:traj-dsm} is a CBF for the augmented system \eqref{eq:sys-aug}. Therefore, there exists $\alpha \in \cK$ such that the DSM-CBF-based safety filter
\begin{subequations}\label{eq:dsm-sf}
  \begin{align}
    \min_{(\bm{u},\bm{w}) \in \sU \times \R^l} \quad &\|\bm{u} - \kappa(\bm{x})\|^2 + \eta \|\bm{w} - \rho(\bm{v})\|^2, \\
    \text{s.t.} \quad & \dot{\Delta}(\bm{x}, \bm{v}, \bm{u}, \bm{w}) \geq - \alpha\big(\Delta(\bm{x}, \bm{v})\big),\label{eq:dsm-cond}
  \end{align}
\end{subequations}
where $\eta > 0$ is small, is feasible for any $(\bm{x}, \bm{v}) \in \tilde{\sC}$.
The function $\rho(\bm{v})$ is called a \emph{navigation field} in the RG literature and plays the same role as $\kappa(\bm{x})$: it is a nominal policy for the virtual reference dynamics $\dot{\bm{v}} = \bm{w}$. Noting
\begin{equation}
  \dot{\Delta} = \frac{\partial \Delta(\bm{x}, \bm{v})}{\partial \bm{x}}\big(f(\bm{x}) + g(\bm{x}) \bm{u}\big) + \frac{\partial \Delta(\bm{x}, \bm{v})}{\partial \bm{v}} \bm{w},
\end{equation}
the safety filter \eqref{eq:dsm-sf} is a quadratic program (QP) if $\sU$ is polyhedral. Unfortunately, \eqref{eq:dsm-sf} is impractical to implement due to two drawbacks: \emph{1)} it can be challenging to solve \eqref{eq:traj-dsm} on an infinite horizon $\tau\in[0,\infty)$ and \emph{2)} differentiability of $\Delta$ is a strong assumption that may not hold in practice. We address these issues in the following.

\subsection{Horizon Considerations}
Since infinite horizons are numerically intractable,
we consider a finite horizon $\tau \in [0,T]$ and employ invariant terminal set constraints to guarantee invariance for $\tau > T$. 
\begin{proposition} \label{cor:fh-traj-dsm}
  Let $T \geq 0$ be the horizon length, and let $\Delta_F:\tilde{\sD} \to \R$ be a DSM with $\tilde{\sC}_F = \{(\bm{x}, \bm{v}) \in \tilde{\sD} \mid \Delta_F(\bm{x}, \bm{v}) \geq 0\}$. If the conditions of Theorem \ref{thm:traj-dsm} hold, then
  \begin{equation} \label{eq:fh-traj-dsm}
    \Delta(\bm{x}, \bm{v}) = \mat{\min_{\tau \in [0,T]} c\big(\Phi(\tau), \bm{v}\big) \\ \Delta_F\big(\Phi(T), \bm{v}\big)}
  \end{equation}
  is a DSM. Moreover, $\tilde{\sC}_F \subset \tilde{\sC} = \{(\bm{x},\bm{v}) \in \tilde{\sD} \mid \Delta(\bm{x}, \bm{v}) \geq 0\}$.
\end{proposition}
\begin{proof}
  Properties \eqref{eq:dsm-safe} and \eqref{eq:dsm-compact} hold by the same arguments made in the proof of Theorem \ref{thm:traj-dsm}. One can similarly show property \eqref{eq:dsm-invariant} for $\Delta_1$.\\
  To prove property \eqref{eq:dsm-invariant} for $\Delta_2$, pick $(\bm{x}, \bm{v}) \in \tilde{\sD}$ such that $\Delta_2(\bm{x}, \bm{v}) = \Delta_F\big(\Phi(T), \bm{v}) = 0$.
  Then,
  \begin{align*}
    \frac{\partial \Delta_2(\bm{x}, \bm{v})}{\partial \bm{x}}f_{\pi}(\bm{x},\bm{v}) &= \frac{\partial \Delta_F\big(\Phi(T), \bm{v}\big)}{\partial \bm{x}} \frac{\partial \Phi(T)}{\partial \bm{x}}  f_{\pi}(\bm{x},\bm{v})\\
    &= \frac{\partial \Delta_F\big(\Phi(T), \bm{v}\big)}{\partial \bm{x}} S_{\bm{x}}(T) f_{\pi}(\bm{x},\bm{v})\\
    &= \frac{\partial \Delta_F\big(\Phi(T), \bm{v}\big)}{\partial \bm{x}} f_{\pi}\big(\Phi(T), \bm{v}\big)\\
    &\geq 0,
  \end{align*}
  where we have used Lemma \ref{lem:sensitivity-property} and property \eqref{eq:dsm-invariant} of $\Delta_F$.

   To prove the final claim $\tilde{\sC}_F \subset \tilde{\sC}$, let $(\bm{x}, \bm{v}) \in \tilde{\sC}_F$. This implies that $\Delta_F(\bm{x}, \bm{v}) \geq 0$ and $(\bm{x}, \bm{v}) \in \tilde{\sD}$.   By continuity of $\Delta_F$ and property \eqref{eq:dsm-invariant}, $\Delta_F\big(\Phi(\tau), \bm{v}\big)\geq 0$ for all $\tau \in [0,\infty)$, implying that $\Delta_2(\bm{x}, \bm{v}) \geq 0$. Moreover, since $\bm{x} \in \sD_{\bm{v}}$, it follows that $\big(\Phi(\tau),\bm{v}\big) \in \tilde{\sC}_F \subset \tilde{\sX}$ for all $\tau \in [0,\infty)$. With this, we conclude that $\Delta_1(\bm{x}, \bm{v}) \geq 0$ and $(\bm{x}, \bm{v}) \in \tilde{\sC}$.
\end{proof}

Suitable options for the terminal DSM $\Delta_F(\bm{x}, \bm{v})$ include the Lyapunov-based DSMs in \cite{freire2026using} or the Sum-of-Squares DSMs in \cite{cotorruelo2021reference}. 
It should also be noted that $\Delta_F(\bm{x}, \bm{v})$ does not need to share the same prestabilizing controller $\pi$ used to compute $\Phi(\tau)$.

\subsection{Smoothness Considerations}
Theorem \ref{thm:traj-dsm} and Proposition \ref{cor:fh-traj-dsm} implicitly assume that, $\forall(\bm{x}, \bm{v}) \in \tilde{\sC}$, the global minimizer of \eqref{eq:traj-dsm} is unique. 
In practice, however, $c(\Phi(\tau),\bm{v})$ may have multiple global minimizers, which causes $\Delta(\bm{x}, \bm{v})$ to not be differentiable. 
The following proposition states that, given a possibly nonsmooth $\Delta$, it is possible to replace \eqref{eq:dsm-cond} with sufficient conditions.

\begin{proposition} \label{prop:nonsmooth-traj-dsm}
  Let $\delta(\tau, \bm{x}, \bm{v}) = c\big(\Phi(\tau), \bm{v}\big)$ and let $\Delta$ be defined as in \eqref{eq:fh-traj-dsm}. For any $T > 0$, if
  \begin{equation} \label{eq:all-tau-cst}
    \dot{\delta}(\tau, \bm{x}, \bm{v}, \bm{u}, \bm{w}) \geq -\alpha\big(\delta(\tau, \bm{x}, \bm{v})\big), \quad \forall \tau \in [0,T],
  \end{equation}
 then, the following inequality holds in the sense of Clarke's generalized Jacobian
  \begin{equation} \label{eq:dsm-cbf-qp-cst}
    \dot{\Delta}_1(\bm{x}, \bm{v}, \bm{u}, \bm{w}) \geq - \alpha\big(\Delta_1(\bm{x}, \bm{v})\big).
  \end{equation}
\end{proposition}
\begin{proof}
  Note that, although $\Delta_1$ may not be differentiable, it satisfies $\Delta_1(\bm{x}, \bm{v}) = \min_{\tau \in [0,T]} \delta(\tau, \bm{x}, \bm{v})$. By compactness of $[0,T]$ and continuity of $\delta$, there exists a set-valued mapping $\tau^*: \tilde{\sD} \to \cP([0,T])$ such that $\forall \tau \in \tau^*(\bm{x}, \bm{v})$, $\Delta_1(\bm{x}, \bm{v}) = \delta(\tau, \bm{x}, \bm{v})$. With this, the Clarke subdifferentials of $\Delta_1$ at $(\bm{x}, \bm{v})$ are the set-valued map
  \begin{equation}
    \sD(\bm{x}, \bm{v}) \!=\! \left\{ \mat{\frac{\partial \delta(\tau, \bm{x}, \bm{v})}{ \partial \bm{x}} & \frac{\partial \delta(\tau, \bm{x}, \bm{v})}{\partial \bm{v}}}\!, \tau \!\in\! \tau^*(\bm{x}, \bm{v}) \right\}\!.
  \end{equation}
  The Clarke's generalized Jacobian of $\Delta_1(\bm{x}, \bm{v})$, denoted $\nabla \Delta_1(\bm{x}, \bm{v})$, is the convex hull of the Clarke subdifferentials.
  We say \eqref{eq:dsm-cbf-qp-cst} holds in the sense of Clarke's generalized Jacobian if, $\forall \mat{H_{\bm{x}} & H_{\bm{v}}} \in \nabla \Delta_1(\bm{x}, \bm{v})$,
  \begin{equation} \label{eq:clarke-condition}
    H_{\bm{x}}\big(f(\bm{x}) + g(\bm{x}) \bm{u} \big) + H_{\bm{v}} \bm{w} \geq - \alpha \big(\Delta_1(\bm{x},\bm{v})\big).
  \end{equation}
   It is clear that \eqref{eq:all-tau-cst} guarantees \eqref{eq:clarke-condition} for all the subdifferentials $\sD(\bm{x}, \bm{v})$ because $\tau^*(\bm{x},\bm{v}) \subset [0,T]$. Since $\nabla \Delta_1(\bm{x},\bm{v})$ is their convex hull, \eqref{eq:clarke-condition} holds for all elements of $\nabla \Delta_1(\bm{x}, \bm{v})$.
\end{proof}

\subsection{Feasibility of the Tractable DSM-CBF Safety Filter}
Incorporating the horizon and smoothness considerations, we arrive at a more tractable formulation of \eqref{eq:dsm-sf}
\begin{subequations}\label{eq:tractable-dsm-cbf-qp}
  \begin{align}
    \min_{(\bm{u},\bm{w}) \in \sU \times \R^l}  &\|\bm{u} - \kappa(\bm{x})\|^2 + \eta \|\bm{w} - \rho(\bm{v})\|^2, \\[3pt]
    \text{s.t.} \qquad & \!\!\!\!\!\!\!\!\dot{\delta}(\tau, \bm{x},\!\bm{v},\! \bm{u},\! \bm{w}) \!\geq\! - \alpha\big(\delta(\tau, \bm{x},\! \bm{v})\big), ~ \forall \tau \in [0,T],\\[5pt]
    & \!\!\!\!\!\!\!\!\dot{\Delta}_F(\Phi(T), \!\bm{v},\! \bm{u},\! \bm{w}) \geq - \alpha\left(\Delta_F\big(\Phi(T), \!\bm{v}\big) \right)\!,\\[-9pt] \nonumber
  \end{align}
\end{subequations}
where $\delta(\tau, \bm{x}, \bm{v}) = c\big(\Phi(\tau), \bm{v}\big)$,
\begin{align*}
  &\dot{\delta}(\tau, \bm{x}, \bm{v}, \bm{u}, \bm{w}) = \frac{\partial c\big(\Phi(\tau),\bm{v}\big)}{\partial \bm{x}} S_{\bm{x}}(\tau) \big(f(\bm{x}) + g(\bm{x}) \bm{u}\big) \\
  &\qquad \qquad \qquad + \frac{\partial c\big(\Phi(\tau), \bm{v}\big)}{\partial \bm{x}} S_{\bm{v}}(\tau) \bm{w} + \frac{\partial c\big(\Phi(\tau), \bm{v}\big)}{\partial \bm{v}} \bm{w},\\
  &\dot{\Delta}_T\big(\!\Phi(\!T\!),\! \bm{v},\! \bm{u},\! \bm{w}\!\big) \!=\! \frac{\partial \Delta_F\big(\Phi(T), \bm{v}\big)}{\partial \bm{x}} S_{\bm{x}}(T) \big(f(\bm{x}) + g(\bm{x}) \bm{u} \big) \\
  & \quad\qquad\quad +\! \frac{\partial \Delta_F\big(\Phi(T), \bm{v}\big)}{\partial \bm{x}} S_{\bm{v}}(T) \bm{w} \!+\! \frac{\partial \Delta_F\big(\Phi(T), \bm{v}\big)}{\partial \bm{v}} \bm{w}.
\end{align*}
While \eqref{eq:tractable-dsm-cbf-qp} has infinite-dimensional constraints, we consider in practice a finite set $\sT \subset [0,T]$ and enforce the constraints $\forall \tau \in \sT$ as outlined in \cite{chen2021backup}. The set $\sT$ can be obtained, along with $\Phi(\tau)$, using a variable-step solver for \eqref{eq:Phi-ivp}. This can be done to an arbitrary degree of accuracy.
As before, \eqref{eq:tractable-dsm-cbf-qp} is a QP if $\sU$ is polyhedral. Since it is natural to question if a CBF-QP retains feasibility for all $(\bm{x}, \bm{v}) \in \tilde{\sC}$, the following corollary proves that  $\bm{u} = \pi(\bm{x}, \bm{v})$ and $\bm{w} = 0$ is always feasible in \eqref{eq:tractable-dsm-cbf-qp}.

\begin{corollary}
  Let $\Delta$ be defined as in \eqref{eq:fh-traj-dsm} with $T \geq 0$. There exists $\alpha \in \cK$ such that the DSM-CBF safety filter \eqref{eq:tractable-dsm-cbf-qp} is feasible for all $(\bm{x}, \bm{v}) \in \tilde{\sC} = \{(\bm{x}, \bm{v}) \in  \tilde{\sD} \mid \Delta(\bm{x}, \bm{v}) \geq 0\}$.
\end{corollary}
\begin{proof}
  Define the set-valued map $\hat{\sC}(c,\tau) \!=\! \{(\bm{x}, \bm{v}) \in \tilde{\sC} \mid 0 \leq \delta(\tau, \bm{x}, \bm{v}) \leq c\}$, and the function
  \begin{equation}
    \hat{\alpha}(c, \tau) = -\inf_{(\bm{x}, \bm{v}) \in \hat{\sC}(c,\tau)} \frac{\partial \delta(\tau, \bm{x}, \bm{v})}{\partial \bm{x}} f_{\pi}(\bm{x}, \bm{v}).
  \end{equation}
  For any $c \geq 0$ and $\tau \in [0,T]$, the set $\hat{\sC}(c,\tau)$ is compact in $\R^n\times \R^l$. It can also be shown that the set-valued map $\hat{\sC}(\cdot, \tau)$ is both upper and lower hemicontinuous at $0$. Thus, it follows by \cite{berge1963topological} maximum theorem that $\forall \tau \in [0,T]$, the function $\hat{\alpha}(\cdot, \tau)$ is continuous at $0$. Moreover, it is nondecreasing because for any $c_1 \leq c_2$, $\hat{\sC}(c_1,\tau) \subset \hat{\sC}(c_2,\tau)$.
  Let $\tau \in [0,T]$ be given and pick any $(\bm{x},\bm{v}) \in \hat{\sC}(0,\tau)$. Since $\Delta_1(\bm{x}, \bm{v}) = \min_{\tau \in [0,T]} \delta(\tau, \bm{x}, \bm{v}) \geq 0$ and $\delta(\tau, \bm{x}, \bm{v}) = 0$, we must have $\frac{\partial \delta(\tau, \bm{x}, \bm{v})}{\partial \bm{x}} f_{\pi}(\bm{x}, \bm{v}) \geq 0$. This implies $\hat{\alpha}(0, \tau) \leq 0$. Therefore, there always exists $\bar{\alpha}(\cdot,\tau) \in \cK$ such that $\forall c \in [0,\infty)$, $\hat{\alpha}(c,\tau) \leq \bar{\alpha}(c,\tau)$. Define $\alpha_1(c) = \max_{\tau \in [0,T]} \bar{\alpha}(c,\tau)$, noting that $\alpha_1(c) \in \cK$. Since $\Delta_F$ is a DSM for $\pi$, it is also a CBF and there exists $\alpha_2 \in \cK$ such that its CBF condition holds.
  Define now $\alpha(c) = \max_{i \in \{1,2\}} \alpha_i(c) \in \cK$. Finally, let $(\bm{x}, \bm{v})\in \tilde{\sC}$ be given and note that $\pi(\bm{x}, \bm{v}) \in \sU$. 
  For any $\tau \in [0,T]$, $\dot{\delta}\big(\tau, \bm{x}, \bm{v}, \pi(\bm{x}, \bm{v}), 0\big) = \frac{\partial \delta(\tau, \bm{x}, \bm{v})}{\partial \bm{x}} f_{\pi}(\bm{x},\bm{v}) \geq -\hat{\alpha}\big(\delta(\tau, \bm{x}, \bm{v}), \tau\big) \geq - \bar{\alpha}\big(\delta(\tau, \bm{x}, \bm{v}), \tau\big) \geq -\alpha_1\big(\delta(\tau, \bm{x},\bm{v})\big) \geq -\alpha\big(\delta(\tau, \bm{x}, \bm{v})\big)$.
  Also, since $\Delta(\bm{x}, \bm{v}) \geq 0$, it follows that $\Delta_F\big(\Phi(T), \bm{v}\big) \geq 0$ and we have that $\big(\pi(\bm{x}, \bm{v}), 0\big) \in \sU \times \R^l$ is such that
  $\dot{\Delta}_T\big(\Phi(T), \bm{v}, \pi(\bm{x}, \bm{v}), 0\big) \!\geq\! -\alpha_2\left(\Delta_F\big(\Phi(T),\bm{v}\big)\right) \geq -\alpha\left(\Delta_F\big(\Phi(T),\bm{v}\big)\right)$.
\end{proof}

Algorithm \ref{alg:dsm-cbf-qp} summarizes the procedure to follow for implementing \eqref{eq:tractable-dsm-cbf-qp} in practice. Note that the input $\bm{u}$ is given to the original system, whereas the virtual input $\bm{w}$ is used to drive the evolution of the virtual reference $\dot{\bm{v}}=\bm w$.\smallskip

\begin{algorithm}[ht] 
  \caption{DSM-CBF Safety Filter}
  \label{alg:dsm-cbf-qp}
  \begin{algorithmic}[1]
  \Require
      \Statex Dynamics $f(\bm{x})$, $g(\bm{x})$
      \Statex Prestabilizing controller $\pi(\bm{x}, \bm{v})$
      \Statex Constraints $c(\bm{x}, \bm{v})$
      \Statex Nominal controllers $\kappa(\bm{x})$, $\rho(\bm{v})$
      \Statex Prediction horizon $T>0$
      \Statex Terminal DSM $\Delta_F(\bm{x}, \bm{v})$
      \Statex Class $\cK$ function $\alpha$
      \Statex 
    \Function{SafetyFilter}{$\bm{x}$, $\bm{v}$}
      \State $\Phi(\tau) \leftarrow$ solve \eqref{eq:Phi-ivp} \Comment $\tau \in [0,T]$
      \State $S_{\bm{x}}(\tau) \leftarrow$ solve \eqref{eq:Sx} \Comment $\tau\in[0,T]$
      \State $S_{\bm{v}}(\tau) \leftarrow$ solve \eqref{eq:Sv} \Comment $\tau\in[0,T]$
      \State $(\bm{u}, \bm{w}) \leftarrow $ solve \eqref{eq:tractable-dsm-cbf-qp}
      \State \textbf{return} $\bm{u}$, $\bm{w}$
    \EndFunction
  \end{algorithmic}
\end{algorithm}

\begin{remark}
    Although \eqref{eq:tractable-dsm-cbf-qp} relies on trajectory predictions, it is important to note that those predictions are always computed for a fixed $(\bm{x}, \bm{v})$. Since the predictions do not depend on the optimization variables $(\bm{u}, \bm{w})$, the computational cost of this approach remains negligible compared to model predictive control.
\end{remark}

\section{Inverted Pendulum on Cart} \label{sec:pend}
\begin{figure}
    \centering
    \includegraphics[width=0.95\linewidth]{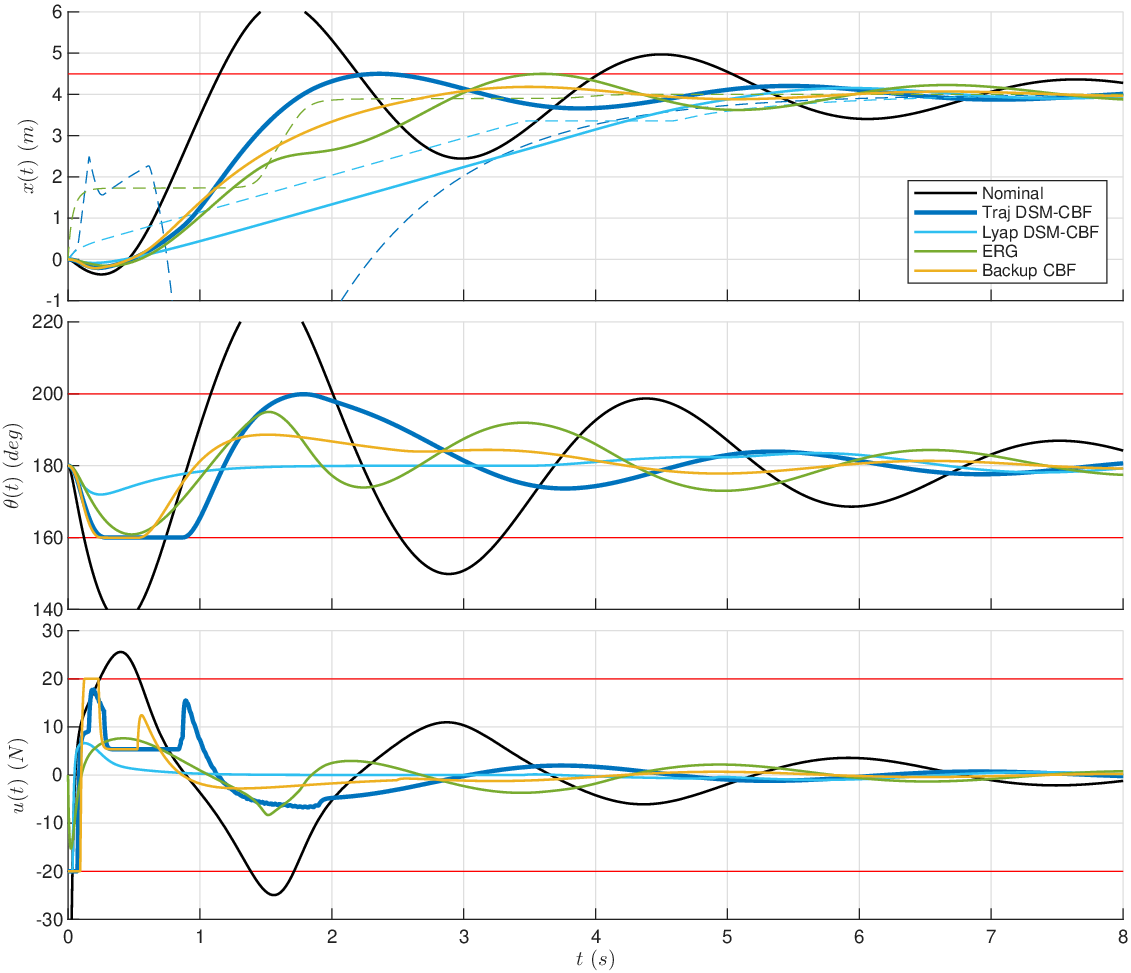}
    \caption{Simulation of the inverted pendulum on a cart under different control approaches. Arguably, the proposed Traj DSM-CBF approach achieves the best performance while guaranteeing constraint satisfaction.}
    \label{fig:pend}
\end{figure}

To showcase the systematic nature of the proposed approach, we apply it to an inverted pendulum on cart with state and input constraints. This example was specifically chosen because it includes several aspects that complicate the design of CBFs, notably: a) input saturation, b) multiple constraints, c) strong nonlinearities, d) high relative degree, and e) non-minimum phase dynamics. The dynamic model of an inverted pendulum on a cart is
\begin{equation}
  M(\bm{q}) \ddot{\bm{q}} + C(\bm{q}, \dot{\bm{q}}) \dot{\bm{q}} + G(\bm{q}) = Bu,
\end{equation}
where the degrees of freedom $\bm{q} = [x; ~ \theta]$ are the cart position $x$ and the pendulum angle $\theta$, and the system matrices are given in \cite{freire2026using}. Letting $\bm{x} = [\bm{q}; ~ \dot{\bm{q}}]$, the system is control-affine and can be written as \eqref{eq:sys}. The equilibrium mapping for this system is $\bar{x}(v) = [v;~ \pi; ~ 0; ~ 0]$ and $\bar{u}(v) = 0$.
We obtain a prestabilizing controller by linearizing the system around the equilibrium point $\bar{x}(v)$, $\bar{u}(v)$ and using an LQR. The resulting controller is $\pi(\bm{x}, v) = -K_\pi \big(\bm{x} - \bar{x}(v)\big)$, with $K_{\pi} = [-0.44~~ 35.3~ -\!1.4~~ 8.0]$.
Consider the task of moving the system from initial position $\bar{x}(0)$ to $\bar{x}(r)$, with $r = 4$, while respecting the state and input constraints
$\sX = \{\bm{x} \in \R^4 : |x| \leq x_{\max}, |\theta - \pi| \leq \theta_{\max}\}$, $\sU = [-u_{\max}, u_{\max}]$, with $x_{\max} = 4.5$, $\theta_{\max} = \pi/9$, and $u_{\max} = 20$. Note that the constraint $|\theta-\pi|\leq\theta_{\max}$ prevents the system from exiting the basin of attraction of the LQR. The reference-dependent state constraint set then becomes $\sX_{\bm{v}} = \{\bm{x} \in \R^4: c(\bm{x}, v) \geq 0\}$, where
$c(\bm{x}, v) = [x_{\max} - x;~ x_{\max} + x;~ \theta_{\max} + \pi - \theta;~ \theta_{\max} - \pi + \theta;~ u_{\max} - \pi(\bm{x},v);~ u_{\max} + \pi(\bm{x}, v)]$. 
The nominal controller $\kappa(\bm{x})$ has the form
$\kappa(\bm{x}, v) = -K_\kappa \big(\bm{x} - \bar{x}(r)\big)$, where $K_{\kappa} = [-35~~ 150 ~ -\!20 ~~ 50]$. The navigation field is $\rho(v) = r - v$. 
We pick a prediction horizon $T= 10$ seconds and construct a Lyapunov-based terminal DSM $\Delta_F$ using the approach in \cite{freire2026using}.
The class $\cK$ function we use for all constraints is $\alpha(c) = 100c$, and $\alpha(c) = 400c$ for the terminal DSM $\Delta_F$. With this, we follow Algorithm \ref{alg:dsm-cbf-qp} to implement the trajectory-based DSM-CBF filter, with $\eta = 0.1$, and its performance is shown in Fig. \ref{fig:pend}.
Also shown are the traces for these approaches:
\begin{enumerate}
  \item \textbf{Nominal}: The nominal performance under the nominal controller $\kappa(\bm{x})$ is unsafe.
  \item \textbf{Lyapunov-based DSM-CBF}: Presented in \cite{freire2026using}, this approach also uses DSMs to design CBFs. However, their reliance on Lyapunov functions makes the approach overly conservative for this open-loop unstable system. 
  \item \textbf{ERG}: Presented in \cite{nicotra2018explicit}, this approach relies on the same trajectory-based DSMs featured in this paper. However, RG approaches are known to be systematically slower than CBFs.
  \item \textbf{Backup CBF}: Presented in \cite{chen2021backup}, this approach can be seen as a precursor to the presented approach since it relies on a ``backup'' policy as opposed to a prestabilizing controller.  Note that there is no systematic way of finding a backup policy. For this comparison, we used $\beta(\bm{x}) = -35.3(\theta - \pi) + 1.4 \dot{x} - 8\dot{\theta}$, where the gains are the same as $K_{\pi}$, but with the first one zeroed-out. This backup policy attempts to keep the pendulum upright and stationary, regardless of its position $x$ (in prestabilizing terms, it is effectively assigning $v(t)=x(t)$). While the approach works well, its response is slower than the proposed method. This is is a direct consequence of the fact that the trajectory-based DSM-CBF uses the reference $v$ to parametrize an entire \emph{family} of backup policies. The added degree of freedom increases performance.
  
\end{enumerate}

\section{Conclusion} \label{sec:conclusion}
This paper presented a systematic approach to design valid CBFs using trajectory predictions of a prestabilized system. Implementation details to achieve a tractable formulation are also provided and studied with rigor. The performance of the proposed approach is illustrated on an inverted pendulum on a cart and compared to other related constrained control approaches. Future work includes exploring different trajectory prediction techniques and studying the robustness properties of the approach under bounded disturbances.

\bibliographystyle{IEEEtran}
\bibliography{references}
\end{document}